\documentclass[a4paper,fleqn,usenatbib]{mnras}

\usepackage[T1]{fontenc}
\usepackage{ae,aecompl}

\usepackage{graphicx}	
\usepackage{amsmath}	
\usepackage{amssymb}	

\newcommand{\pegase}{\textsc{P\'egase}}
\newcommand{\herschel}{\textit{Herschel}}
\newcommand{\pacs}{\textsc{pacs}}
\newcommand{\spire}{\textsc{spire}}
\newcommand{\spitzer}{\textit{Spitzer}}

\title[Starbursts and dusty tori in distant 3CR radio galaxies]{Starbursts and dusty tori in distant 3CR radio galaxies}

\author[P. Podigachoski et al.]{
Pece Podigachoski,$^{1}$\thanks{E-mail: podigachoski@astro.rug.nl}
Brigitte Rocca-Volmerange,$^{2,3}$
Peter Barthel,$^{1}$\newauthor
Guillaume Drouart,$^{4}$
and Michel Fioc$^{2}$
\\\\
$^{1}$Kapteyn Astronomical Institute, University of Groningen, PO Box 800, NL-9700 AV Groningen, the Netherlands\\
$^{2}$Institut d'Astrophysique de Paris, Universite Pierre et Marie Curie/CNRS, 98 bis Bd Arago, F-75014 Paris, France\\
$^{3}$Universite Paris-SUD, F-91405 Orsay Cedex, France\\
$^{4}$International Centre for Radio Astronomy Research, Curtin University, Bentley, WA6102, Perth, Australia
}

\date{Accepted XXX. Received YYY; in original form ZZZ}

\pubyear{2016}

\begin{document}
\label{firstpage}
\pagerange{\pageref{firstpage}--\pageref{lastpage}}
\maketitle

\begin{abstract}
We present a study of the complete ultraviolet to submillimetre spectral energy distributions (SEDs) of twelve 3CR radio galaxy hosts in the redshift range $1.0 < z < 2.5$, which were all detected in the far-infrared by the \textit{Herschel Space Observatory}. The study employs the new spectro-chemical evolutionary code {\pegase}.3, in combination with recently published clumpy AGN torus models. We uncover the properties of the massive host galaxy stellar populations, the AGN torus luminosities, and the properties of the recent starbursts, which had earlier been inferred in these objects from their infrared SEDs. The {\pegase}.3 fitting yields very luminous (up to 10$^{13}$\,L$_{\odot}$) young stellar populations with ages of several hundred million years in hosts with masses exceeding 10$^{11}$\,M$_{\odot}$. Dust masses are seen to increase with redshift, and a surprising correlation -- or better upper envelope behaviour -- is found between the AGN torus luminosity and the starburst luminosity, as revealed by their associated dust components. The latter consistently exceeds the former by a constant factor, over a range of one order of magnitude in both quantities. 
\end{abstract}



\section{Introduction}
The episodic accretion of matter onto supermassive black holes (SMBH), which are nowadays assumed to exist in the central regions of almost all massive galaxies in the Universe, results in phenomena known as active galactic nuclei (AGN). Because of the interaction with their host galaxies, primarily exhibited via negative and/or positive feedback processes, AGN are essential elements -- or phases -- in the evolution of galaxies through cosmic time. Particular attention in studies of galaxy evolution is paid to the early epoch $1<z<3$, which is the time when both the stellar bulges of galaxies and their associated black holes went through peak growth \citep{Alexander&Hickox12,Heckman&Best14}.  

A small fraction of AGN are characterized by strong radio emission, which is produced by the powerful radio-jets and radio-lobes driven by the growth of the black hole. The most powerful radio-loud AGN are the so-called FRII sources \citep{Fanaroff&Riley74}. By virtue of their huge radio luminosities, these sources were historically used in searches for the most distant objects in the Universe \citep{Roettgering94,Stern&Spinrad99}. While no longer being the highest redshift holders, radio-loud AGN (and their hosts) have remained central in studies of the interplay between AGN and star formation (SF) activity in massive galaxies in the early Universe \citep[see][for a review]{Miley&DeBreuck08}. The reasons are straightforward: high-$z$ radio-loud AGN are powerful AGN \citep{Haas08,DeBreuck10,Dicken14}, and their hosts are among the most massive galaxies in the Universe \citep{Best98,Seymour07} often showing prodigious levels of SF activity \citep{Archibald01,Reuland04,Drouart14,Tadhunter14,Podigachoski15a}. From the sharp cut in the Hubble K-band diagram, and using elliptical scenarios corrected for evolution and cosmology, \citet{RoccaVolmerange04} estimated a maximum stellar mass of 10$^{12}$\,M$_{\odot}$ for $1 < z < 4$ objects. 

Additionally, radio-loud AGN are often used in unification studies. Within the framework of the unified model of radio-loud AGN \citep{Barthel89}, FRII radio galaxies (RGs, type 2) and radio-loud quasars (type 1) are assumed to belong to the same parent population, and can be unified based on orientation \citep[see][for a recent review]{Antonucci12}. Central to this model is the AGN torus, a region rich in molecular gas and dust perpendicular to the radio source axis, obscuring the accretion disc and the broad-line region along a substantial range of viewing angles \citep[e.g.,][]{Drouart12}. In the case of RGs, the torus acts like a natural coronograph, blocking most of the UV/optical light emitted due to the AGN activity \citep[e.g.,][]{Wilkes13} and -- in contrast to quasars -- enabling detailed studies of the stellar populations of the AGN host galaxy. 

Originally selected at low radio frequencies (178~MHz), the landmark Revised Third Cambridge Catalogue of Radio Sources \citep[hereafter 3CR;][]{Spinrad85} contains some of the brightest radio-loud AGN at all redshifts. The $z>1$ double-lobed (FRII) RGs and quasars in the 3CR sample almost universally accrete at high Eddington rates, i.e. in quasar-mode \citep[e.g.,][]{Best&Heckman12}, resulting in an unbiased sample, free of any of the low-power AGN often found at lower redshifts ($z\sim0.5$). The $z>1$ part of this sample has been observed with virtually all space-based telescopes \citep{Best97,Haas08,Leipski10,Wilkes13,Chiaberge15}, including the {\herschel} telescope \citep{Barthel12,Podigachoski15a}. 

Using primarily {\spitzer} and {\herschel} broad-band photometry, \citet{Podigachoski15a} decomposed the rest-frame infrared (IR) spectral energy distributions (SEDs) of the complete $z>1$ 3CR sample into AGN- and SF-related components, adopting for the latter a typical modified blackbody with a fixed dust emissivity index. Compared to studies of other radio-loud AGN in the high-$z$ Universe \citep[e.g.,][]{Drouart14}, \citet{Podigachoski15a} found a somewhat higher {\herschel} detection fraction, with about half of their sample objects detected in at least three {\herschel} bands, indicating rates (SFRs) of several hundred solar masses per year. Such prodigious SFRs, at the level of those of typical submillimetre galaxies at similar redshifts, were found despite the powerful AGN activity which often dominates the IR luminosities of the 3CR objects, ruling out uniform quenching of star formation and providing tentative evidence for jet-triggering of star formation \citep{Podigachoski15a}. Adopting a modified blackbody component to account for the emission of the star-formation-heated dust provides robust estimates of the temperature and the mass of this cold dust in AGN hosts; however, it yields no information on the mass, age, and/or metallicity of the (young) stellar component which powers the dust emission in the rest-frame far-infrared (FIR). Such information, particularly for AGN hosts which are not completely enshrouded by dust, can be robustly obtained by also considering the unattenuated part of the stellar continuum due to the young stars. Furthermore, by only exploring SEDs beyond 1~$\mu$m in the rest-frame of the objects, \citet{Podigachoski15a} do not consider the evolved stellar populations in the hosts of 3CR AGN, which are expected to peak at about 1~$\mu$m in the rest-frame. 

Here, we extend the work of \citet{Podigachoski15a} by studying the rest-frame ultraviolet (UV) to submillimetre (submm) SEDs of twelve $z>1$ 3CR RG hosts, identifying the emissions from the past and recent stellar populations and from the AGN torus, with the goal to constrain the physical properties of these stellar populations. We use the new spectro-chemical evolutionary code {\pegase}.3 (Fioc \& Rocca-Volmerange, in preparation), which follows the masses of stars, gas and dust, determines the attenuation of stellar emission by grains in H\,\textsc{ii} regions and the diffuse interstellar medium (ISM), and consistently computes the emission of the latter. Galaxy templates computed with {\pegase}.3 were earlier used by \citet{RoccaVolmerange13}, who performed a pilot study of two distant ($z=3.8$) RGs selected for their small AGN contributions. Coupling {\pegase}.3 templates with smooth torus models by \citet{Fritz06}, \citet{Drouart16} recently extended this pilot study to a sample of 11 more powerful RGs from the $1 < z < 4$ sample of \citet{Drouart14}. Here, we use the latest torus models in the literature \citep{Siebenmorgen15} in a study of a dozen well-known 3CR RGs, also aiming to test the different AGN torus formalisms. We consider only RGs, but maintain the view that most results obtained for this AGN class are applicable also for radio-loud quasars within the unified model of radio-loud AGN \citep[e.g.,][]{Podigachoski15b}.    

This paper is organized as follows. In \S~\ref{sec:data} we present the sample used in this work, and the observational data which we use as input for our SED fitting approach. In \S~\ref{sec:templates} we provide overview of the {\pegase}.3 model predictions including a coherent dust emission, and of the adopted library of torus models. The results, including the best-fit SEDs of each sample object, are presented in \S~\ref{sec:results}, and discussed in \S~\ref{sec:discussion}. We conclude this paper with a brief summary (\S~\ref{sec:conclusions}). The Appendix contains details on the observational data for each object studied in this work.  
\section{Sample selection and data}
\label{sec:data}
\begin{figure}
\includegraphics[width=\columnwidth]{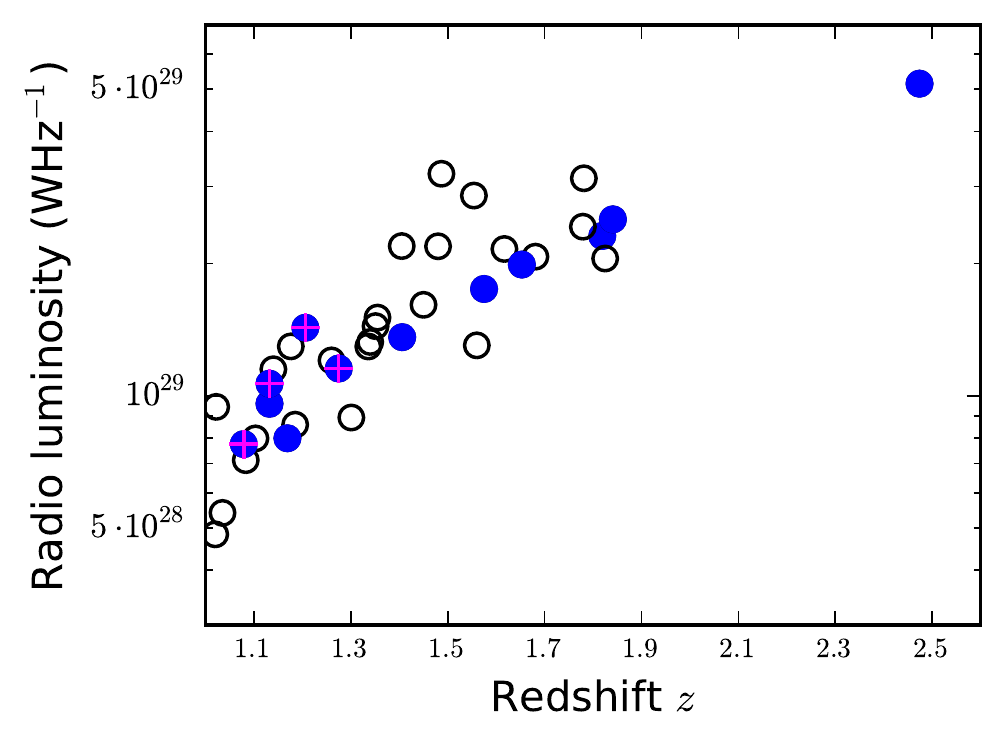}
\caption{Sample selection. Plotted is the observed radio (178 MHz) luminosity as a function of redshift for the complete $z>1$ 3CR sample of radio galaxies. The subsample of radio galaxies with good signal-to-noise detections from UV to submm wavelengths, which is the one considered in this work (see text), is shown with filled symbols. Objects belonging to this subsample for which {\spitzer} spectra are available, are indicated with crosses.}
\label{fig:selection}
\end{figure}
The RGs studied in this work belong to the Third Cambridge Catalog of Radio Sources \citep{Bennett62,Spinrad85}. Being selected due to their steep-spectrum radio emission from their radio-lobes, these objects are some of the most powerful AGN at any redshift in the Universe. The sample studied in this work is a subset of the complete high-$z$ ($z>1$) catalog of 3CR RGs, which contains a total of 37 RGs \citep{Spinrad85}. The highest redshift object is 3C~257 at $z=2.474$, while all other sources are at redshifts $1<z<2$.  
\begin{table}
\centering
\caption{The twelve objects selected from the complete $z>1$ 3CR sample of radio galaxies and studied in this work. The star symbols next to the names indicate objects for which {\spitzer} IRS spectra are available.}
\label{tab:sample}
\begin{tabular}{lcc}
\hline
Name & $z$ & log(L$_{\mathrm{178~MHz}} (\mathrm{W~Hz^{-1}})$) \\
\hline
3C~068.2 & 1.57 & 29.2 \\
3C~210 & 1.17 & 28.9 \\
3C~256 & 1.82 & 29.4 \\
3C~257 & 2.47 & 29.7 \\
3C~266* & 1.27 & 29.1 \\
3C~297 & 1.41 & 29.1 \\
3C~305.1 & 1.13 & 29.0 \\
3C~324* & 1.21 & 29.2 \\
3C~356* & 1.08 & 28.9 \\
3C~368* & 1.13 & 29.0 \\
3C~454.1 & 1.84 & 29.4 \\
3C~470 & 1.65 & 29.3 \\
\hline
\end{tabular}
\end{table}

Decomposing the spectral energy distributions of RGs into stellar and AGN-related components requires observations at the MIR. The complete high-$z$ 3CR sample has been observed at six different bands at wavelengths between 3.6 and 24~$\mu m$ with all {\spitzer} imaging instruments \citep{Haas08}. Given that almost all RGs have been detected with all {\spitzer} instruments at good signal-to-noise ratios, the {\spitzer} photometry does not introduce any selection effects. We note however, that such effects are introduced when selecting objects which have been detected in the rest-frame FIR and UV/optical bands, as described below.        

A crucial step forward in understanding the properties of the high-$z$ 3CR sample has recently been provided by {\herschel} imaging, using both imaging instruments ({\pacs} at 70 and 160, and {\spire} at 250, 350, and 500 $\mu m$). Full details on the data reduction and photometry of the 3CR sample are provided by \citet{Podigachoski15a}. As shown in that paper, the typical temperature of the cold dust in 3CR hosts is about 40~K, which means that the peak due to the dust-reprocessed young stellar emission occurs at around 70 $\mu m$. To constrain this peak, we require that each RG is detected in at least one photometric band at rest-frame wavelength greater than 70 $\mu m$, which given the redshift of our RGs requires that the objects are detected in the {\spire} 250 $\mu m$ band\footnote{Note that all objects with a robust {\spire} 250 detection are also detected in both {\pacs} bands.}. This is in fact the main criterion we impose when selecting the RGs for the current work, and this criterion is only relaxed in two cases, 3C~210 and 3C~356, because despite them being detected in only the two {\pacs} bands, their sufficiently low redshift ensures that the {\pacs} 160 $\mu m$ band probes emission beyond the infrared peak. Fifteen RGs satisfy this selection criterion. Clearly, selecting the {\herschel}-detected RGs means that our work features only the most prodigiously star-forming 3CR RGs. However, as pointed out by \citet{Podigachoski15a}, the objects not detected with {\herschel} might still actively form stars, though at a significantly lower level.   

In addition to powering the dust-reprocessed emission in the FIR, young stars -- when not attenuated -- also emit strongly in the UV/optical. Furthermore, the evolved stellar populations produce stellar continuum radiation which peaks in the optical/NIR \citep{RoccaVolmerange13}. Both considerations render UV-to-NIR observations crucial to our SED study of RGs. A major study of the properties of $z\sim1$ 3CR RGs was performed by \citet{Best97}, whose work is the source of most NIR data and about half of the UV/optical data used in our work\footnote{The \citet{Best97} \textit{Hubble Space Telescope} observations were later re-analysed by \citet{Inskip06}, who extracted the photometry of the 3CR RGs in 4{\arcsec} apertures as opposed to the 9{\arcsec} (diameter) apertures used by \citet{Best97}. To compute colors at the exact same filters for each RG, in most cases \citet{Inskip06} relied on interpolations based on observations in other nearby filters. To avoid the additional uncertainty associated with this step, we choose the \citet{Best97} measurements despite the fact that some of our objects may have some contamination from nearby objects within the larger aperture (see \S~\ref{sec:limitations}).}. More UV/optical observations with the improved \textit{Hubble Space Telescope (HST)} were recently obtained as part of an \textit{HST SNAPSHOT} programme \citep{Chiaberge15, Hilbert16}. To better constrain the ages of the stellar populations, we require that each RG has at least two photometric observations, one on either side of the 4000{\AA} break. This selection criterion limits the 15 {\herschel}-detected objects to 13, which can be grouped as follows: 8 have at least two \textit{HST} and one NIR observation \citep[mainly provided by][]{Best97}, and 5 have only two \textit{HST} observations \citep[mainly provided by][]{Chiaberge15} and no NIR observations. The prominent emission lines seen in the UV/optical spectra of high-$z$ RGs \citep[e.g.,][]{McCarthy93,Best00}, including but not limited to \ion{C}{II}, \ion{Ne}{V}, \ion{O}{II}, \ion{H}{$\beta$}, \ion{O}{III} and \ion{H}{$\alpha$} may have an important contribution ($\sim$30\%) to the total flux measured in broadband filters; these line fluxes have been subtracted based on optical spectra as explained in the corresponding reference papers \citep{Best97,Hilbert16}.   

Hence, the final sample studied in this work contains 12 well-known type 2 AGN of the 13 RGs which satisfy both selection criteria: we remove 3C~119 from the subsequent analysis, because it has been shown that it is an object with a quasar-like NIR/MIR SED \citep[see][]{Podigachoski15b}. The location of the sample in the radio luminosity-redshift plane is shown in Fig.~\ref{fig:selection}. As shown in that figure, the selected RGs are homogeneously distributed in redshift. The main parameters of the final sample addressed in this work are summarized in Table~\ref{tab:sample}. Tables listing the details of the available photometry used for the fitting for each of these RGs are presented in the Appendix.  

In addition to the photometric observations presented above, some\footnote{\citet{Leipski10} in fact studied the complete sample of 3CRR radio-loud AGN at the redshift range $1<z<1.4$, but because the 3CRR catalogue is a subsample of the parent 3CR catalogue studied in this work, the few 3CR RGs in this redshift range not belonging to 3CRR do not have {\spitzer} spectroscopy.} 3CR RGs with redshifts $z<1.4$ have been observed in spectroscopic mode with the IRS spectrograph on {\spitzer} \citep{Leipski10}. The availability of such spectra is not a selection criterion in our work: the spectra are merely used to confirm that the torus models which we apply (see below) account well for the MIR photometry of RGs, and that the disentangling of AGN and young stellar components in the important MIR/FIR transition region is reasonably robust. Given the spectral window covered by the IRS spectrograph, between 19.5 and 36.5 $\mu$m, and the redshift range of the 3CR RGs given above, the {\spitzer} spectra often contain both low- and high-excitation MIR emission lines (e.g., \ion{Ne}{II}, \ion{Ne}{V}, \ion{Ne}{III}), the 11.3 $\mu$m line from PAH features, and the 9.7 $\mu$m silicate feature \citep{Leipski10}. To include the spectroscopic information to the SED analysis, we measure flux densities at three artificial broadband filters centred at 27, 30, and 33 $\mu$m. The RGs which have {\spitzer} spectra are: 3C~266, 3C~324, 3C~356, and 3C~368 (see Table~\ref{tab:sample}).
 
Finally, some objects from our final sample also have SCUBA/MAMBO submm (850 $\mu m$) observations. Given that our high-$z$ 3CR RGs are exclusively steep-spectrum radio sources, any synchrotron emission from the radio lobes can safely be neglected, and in practice, these observations probe the Rayleigh-Jeans tails of the thermal dust continuum emission \citep[see e.g.,][]{Haas06,Podigachoski15a}.  
\section{Models and templates}
\label{sec:templates}
\begin{figure*}
\includegraphics{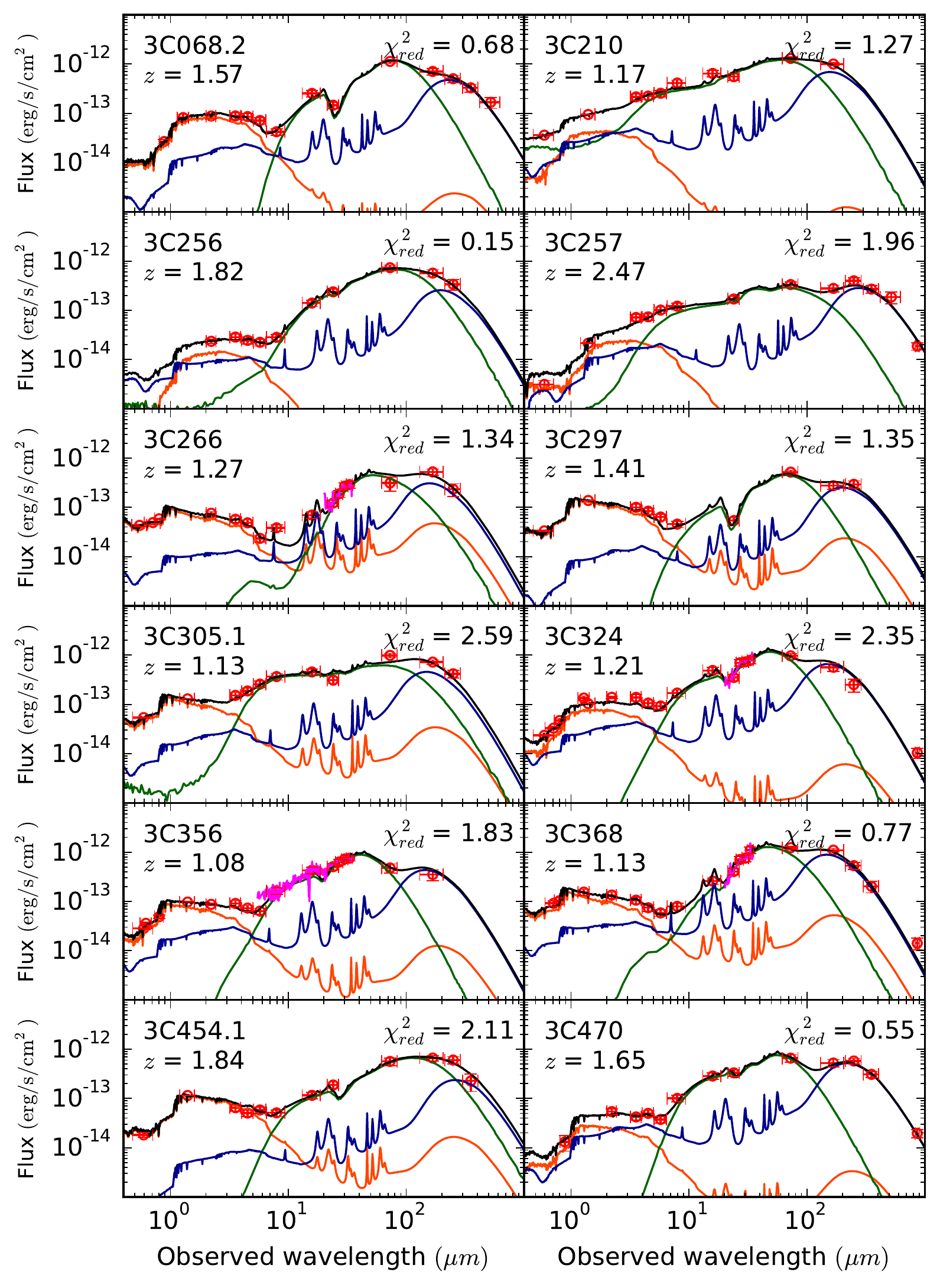}
\caption{The best-fit UV-to-submm spectral energy distributions (SEDs) of the 3CR radio galaxies studied in this work. The observed SEDs (in units $\lambda F_\lambda$) are shown with red symbols. The components combining to yield the total SED (black) are as follows: old stellar component modelled with {\pegase}.3 (orange), young stellar component modelled with {\pegase}.3 (blue), and AGN-powered torus component (green) from the library of \citet{Siebenmorgen15}. Overplotted in magenta are the {\spitzer} IRS spectroscopic observations, available for selected sources. Zoom-ins of the regions probed by these spectra are shown in Fig.~\ref{fig:irs}.}
\label{fig:seds}
\end{figure*}
The templates used in this work are built with the help of two models: the evolutionary synthesis code {\pegase}.3 (Fioc \& Rocca-Volmerange, in preparation), and the AGN torus model by \citet{Siebenmorgen15}. Both models use Monte Carlo simulations for solving the radiative transfer equations. 
\subsection{{\pegase}.3 models}
{\pegase}.3, predicts the attenuated star and gas emission, and the coherent dust emission from the attenuation, coherently transferred through the diffuse ISM and star forming regions. Built as a flexible tool, {\pegase}.3 provides an extensive range of choices in terms of star formation rates, initial mass functions, and infall or outflow rates to mimic the galaxy evolution for starbursts and various spectral types over a large redshift range. It tracks the chemical ISM enrichment of specifically the elements C and Si, plus O, Fe, N, and other metals, from which dust mass grows. At any wavelength, the code sums the contributions of the stellar photospheric emission, nebular continuum and the dust emission of the heated grains. Destruction and accretion dust phases are considered using classical Draine models and specific modelling \citep{Dwek98}. Through Monte Carlo simulations of radiative transfer, the attenuation (absorption and scattering) and dust re-emission (including stochastic heating) are predicted for various geometries (slab, spheroids, bulges + disks). The main difference of {\pegase}.3 with most other models is the coherence in metal enrichment from the stellar ejecta derived from the adopted star formation law. Extended to a large coverage of wavelength domain and respecting energy conservation, it is used in place of extinction laws or any UV-far-IR relations. Galaxy scenarios by type are defined with a limited number of free parameters (star formation rate, infall, outflow and initial mass function) to be compatible with the local SDSS galaxy color-color distribution \citep{Tsalmantza12}. The detailed documentation (at \url{www2.iap.fr/pegase}) provides the possibility to propose new scenarios among which those derived from semi-analytic models or numerical simulations. As 3CR RG hosts are expected to be massive ellipticals occasionally undergoing bursts of star formation, we consider here only starburst and early type scenarios, as used by \citet{RoccaVolmerange13} and \citet{Drouart16}. 
\subsection{AGN torus models}
The high-$z$ 3CR RGs are demonstrably powerful emitters in the rest-frame MIR \citep{Haas08,Podigachoski15a}, where thermal emission from the AGN torus dust is the most dominant emission process. The physical origin of and the distribution of dust within the AGN torus are hotly debated issues in the literature. The two main families of models in the literature are the smooth models \citep[e.g.,][]{Fritz06}, whereby dust is distributed homogeneously throughout the torus, and the clumpy models \citep[e.g.,][]{Nenkova08,Hoenig&Kishimoto10}, whereby dust is located in individual dusty clouds filling the torus. A number of photometric and spectroscopic studies have been performed aiming to pin down the relevant physical parameters within a given formalism, yet the best evidence (albeit from a small number of nearby objects) comes from MIR interferometric studies \citep[e.g.,][and references therein]{Hoenig13}. 

In this work, we adopt the models presented by \citet{Siebenmorgen15}, whose formalism includes both a homogeneous disc of gas and individual dust clouds randomly distributed throughout the torus. The five free parameters of these models are: (i) viewing angle (9 values), (ii) inner torus radius (5 values), (iii) volume filling factor of clouds (4 values), (iv) optical depth of clouds (4 values), and (v) optical depth of disc midplane (5 values), which results in a library of 3600 unique torus models. Given the dense environment in circumnuclear regions, the \citet{Siebenmorgen15} models adopt fluffy dust grains, a choice resulting in stronger FIR and submm emission compared to other clumpy models, but also a more pronounced NIR emission because the larger (than standard ISM) fluffy grains can survive much closer to the AGN. This, together with the additional material in the innermost region of the torus provided by the homogeneous disc and the dust clouds in the ionization cones of the AGN naturally reproduces the NIR bump seen in some RGs. Note that given the limited photometric data in the MIR in this work, we cannot study the torus in detail, and we stress that our goal is to simply isolate the AGN-powered emission from the stellar emission.  
\subsection{SED fitting procedure}
\label{sec:fitting}
Following the approach by \citet{Drouart16}, synthetic libraries of early-type galaxy and starburst plus AGN templates are built with a variety of parameters over large ranges. 
The free parameters for the early-type galaxy templates include the normalization and the age of the stellar component, and the ones for the starburst templates include the initial metallicitiy in addition to the age and normalization of this stellar component. More than 10$^7$ templates are tested for comparison to the global observed spectral energy distributions. The instantaneous starburst \citep[see][]{RoccaVolmerange13} is preferred to test the most powerful emitters in the far-IR: its parameters are a \citet{Kroupa93} initial mass function, and no outflow or infall. As described by \citet{Drouart16}, the AGN templates are added to each starburst template as a relative contribution at 20 $\mu$m, using a grid of values ranging between dominant to negligible AGN contribution at this particular wavelength. Observations are compared to the sum of normalized SED templates by a $\chi^2$-minimization procedure \citep{LeBorgne&RoccaVolmerange02} on the largest wavelength coverage, providing the best-fit of the bolometric luminosity. The calibration factor is derived from the global best-fit synthetic SED compared to the observations, which is then applied to all normalized {\pegase}.3 outputs (e.g., masses, luminosities, etc.)
\section{Results}
\label{sec:results}
\begin{figure}
\includegraphics[width=\columnwidth]{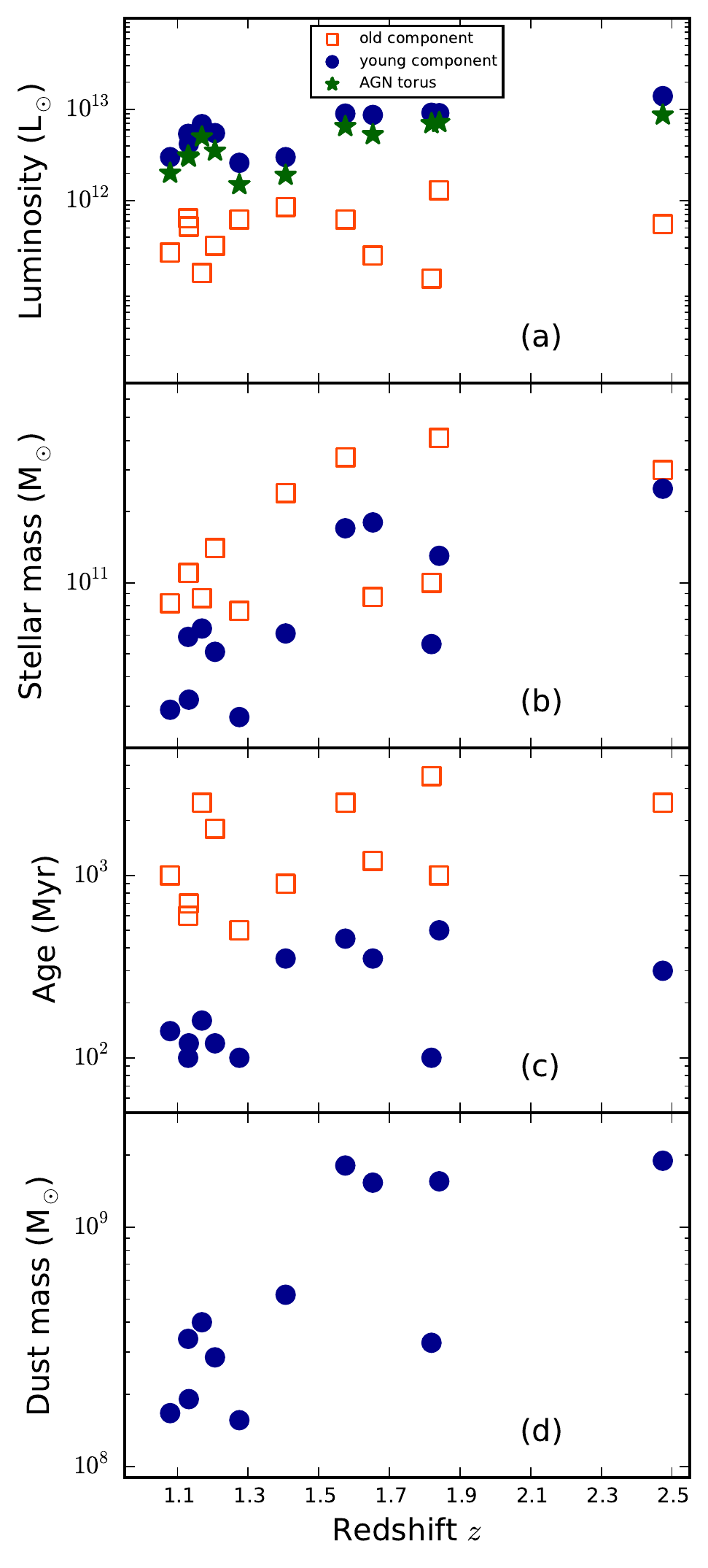}
\caption{Various physical properties of the 3CR radio galaxies obtained using the fitting approach adopted in this work, as a function of redshift. Results for the old and young stellar component are plotted as (orange) empty squares and (blue) circles, respectively, and the AGN-powered torus component is plotted as (green) stars. (a) Luminosities obtained by integrating the relevant best-fit components as described in the text. (b) Masses of the best-fit stellar components. (c) Ages of the best-fit stellar components. (d) Dust mass produced during the evolution of the young stellar component.}
\label{fig:all_vs_z}
\end{figure}

In this section we discuss the overall successes and limitations of the SED fitting approach, and the results obtained from the best-fit SEDs. The best-fit SEDs are presented in Fig.~\ref{fig:seds} and Fig.~\ref{fig:irs}, and details on the observations of twelve individual RG hosts are provided in the Appendix. We mainly focus on the physical properties of the young stellar component obtained from the best-fits, and the possible link between its infrared luminosity and that of the AGN torus component. In addition to the infrared luminosity, which is computed by integrating the young stellar component's SED from 1 to 1000 $\mu$m, these properties include the age and mass of young stars, and the mass of dust which has been produced during the evolution of the starburst. We also report the stellar masses, ages, and the bolometric luminosities (integrating from 0.09 to 1000 $\mu$m) of the evolved stellar component. All results are tabulated in Table~\ref{tab:results}, and some are plotted in Fig.~\ref{fig:all_vs_z}.  
\subsection{Spectral energy distributions}
\label{sec:seds}
\begin{figure}
\includegraphics[width=\columnwidth]{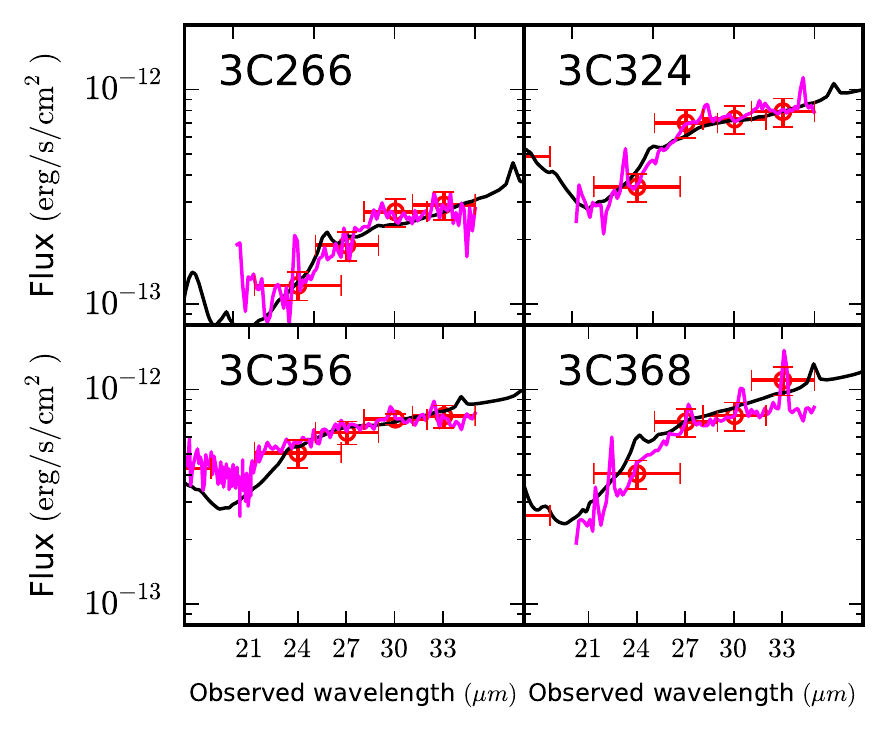}
\caption{Spectral energy distributions (SEDs) in the wavelength domain probed by the {\spitzer} IRS spectra, for the four objects from our sample with such spectra available \citep{Leipski10}. The IRS spectra and the total SEDs (as presented in Fig.~\ref{fig:seds}) are plotted in magenta and black colours, respectively. The photometric data in this wavelength domain (see relevant tables in Appendix) are indicated with red symbols. 
Note the  deep silicate feature in the spectrum of 3C~356 (see text).}
\label{fig:irs}
\end{figure}
As shown by \citet{RoccaVolmerange13,RoccaVolmerange15} and by \citet{Drouart16}, the presence of different stellar populations in RG hosts and their strongly accreting supermassive black holes necessitate a fitting approach involving three spectral components, namely the old stellar component, the young stellar component with associated cold dust\footnote{In the case of 3C~305.1 and 3C~454.1, the young stellar component practically does not contribute to the UV/optical photometry and the {\herschel} photometry could be consistent with the long-wavelength extension of the torus. Hence, in these cases, the sum of merely an old stellar component and an AGN torus component might provide a good representation of the observed SEDs.}, and the warm AGN torus dust component reemiting the intense AGN radiation. On the one hand, as shown in Fig.~\ref{fig:seds}, some of these components can completely dominate the emission in some wavelength domains, as exemplified by the AGN torus emission in the MIR. In other wavelength domains, on the other hand, often two (or sometimes all three) components are needed to successfully reproduce the observed SEDs. The AGN-powered torus emission peaks between 20 and 40 $\mu$m in the rest-frame MIR, and the young-star-powered dust grain emission peaks between 65 and 90 $\mu$m (see Fig.~\ref{fig:seds}). With reference to Fig.~\ref{fig:irs}, we note that in all cases where {\spitzer} IRS spectra are available, they are exceptionally well reproduced by the torus models, strengthening the view that the disentangling of the torus and cold dust grain emission is done in a robust manner.        

The SEDs of the high-$z$ 3CR RGs show a number of characteristic features which we discuss below. The IRS spectra of objects in our work \citep[but also of other 3CR objects from the complete sample, see][]{Leipski10} show that PAH features are not prominent \citep[but see][]{Rawlings13}, suggesting that the MIR emission of the high-z 3CR RGs is completely dominated by the AGN activity (see Fig.~\ref{fig:irs}), which is also what is consistently seen in the best-fit models for objects which do not have MIR spectroscopic observations (Fig.~\ref{fig:seds}). Moreover, the SEDs of some RGs (e.g., 3C~068.2, 3C~324, 3C~356) show deep silicate features at 9.7 $\mu$m rest-frame, and in the case of 3C~356, this silicate feature is also detected in the IRS spectrum shown in Fig.~\ref{fig:seds} and Fig.~\ref{fig:irs} (at an observed wavelength of 20.2 $\mu$m). The NIR bump seen in the observed SEDs of some 3CR RGs \citep[e.g.,][]{Podigachoski15a} is also well reproduced with the \citet{Siebenmorgen15} library of torus models. In some cases, the strong AGN activity responsible for this NIR bump can outshine the 1 $\mu$m peak usually associated with the evolved stellar population in RGs (e.g., 3C~257).

While the adopted three-component approach successfully reproduces the observed SEDs of all RGs, it may not give satisfactory physical results for some particular objects. This mainly concerns the estimated ages of the old stellar components associated with 3C~256, 3C~266, and 3C~368, which as discussed below are estimated to be the youngest throughout the sample. As argued by \citet[][for 3C~266 and 3C~368]{Best98} and \citet[][for 3C~256]{Simpson99}, these sources are among the ones with the strongest alignment effects\footnote{This is the well-known co-spatial extent of the radio and UV/optical emission, resulting from the interaction between the radio-jet and the interstellar matter of the host galaxy in combination with the scattered AGN light \citep{Chambers87,McCarthy87}.}, and as such we suspect a significant non-stellar contribution to their UV/optical SEDs. However, including yet another spectral component to the radio galaxies' SEDs, or treating some of the UV/optical flux as contributed by direct AGN emission and/or scattered AGN light is outside the scope of this work. Nevertheless, the results obtained for these sources (as indicated below) should be treated with care \citep[see for instance][for a more detailed analysis of the optical SED of 3C~256]{Simpson99}. The case for a non-stellar contribution is further discussed in \S~\ref{sec:limitations}.     
\begin{table*}
\centering
\caption{Best-fit physical parameters obtained from spectral energy distribution fitting. The two fits for 3C~256 obtained with and without the UV/optical photometric data (see \S~\ref{sec:limitations}), are denoted with superscripts (a) and (b), respectively. Columns represent the following: (1) Object name; (2) bolometric luminosity of the evolved (old) stellar component; (3) infrared (1 to 1000 $\mu$m) luminosity of the young (starbursting) stellar component; (4) infrared (1 to 1000 $\mu$m) luminosity of the AGN torus component; (5) stellar mass of the old stellar component; (6) stellar mass of the young stellar component; (7) dust mass of the young stellar component; (8) age of the old stellar component; (9) age of the young stellar component; (10) metallicity of the old stellar component; and (11) initial metallicity of the young stellar component.}
\label{tab:results}
\begin{tabular}{lcccccccccc}
\hline
Name & L$^{\mathrm{old}}$ & L$^{\mathrm{young}}$ & L$^{\mathrm{torus}}$ & M$_{\mathrm{stellar}}^{\mathrm{old}}$ & M$_{\mathrm{stellar}}^{\mathrm{young}}$ & M$_{\mathrm{dust}}^{\mathrm{young}}$ & Age$^{\mathrm{old}}$ & Age$^{\mathrm{young}}$ & Z$\ast^{\mathrm{old}}$ & Z$\ast^{\mathrm{young}}$ \\
& $10^{11}$ [L$_{\odot}$] & $10^{12}$ [L$_{\odot}$] & $10^{12}$ [L$_{\odot}$] & $10^{11}$ [M$_{\odot}$] & $10^{11}$ [M$_{\odot}$] & $10^{9}$ [M$_{\odot}$] & [Myr] & [Myr] & \\
\hline
3C~068.2 & 6.2 & 9.0 & 6.5 & 3.4 & 1.7 & 1.8 & 2500 & 450 & 0.0187 & 0.0010 \\
3C~210 & 1.6 & 6.9 & 5.0 & 0.9 & 0.6 & 0.4 & 2500 & 160 & 0.0187 & 0.0010 \\
3C~256$^{(a)}$ & 5.9 & 8.8 & 6.6 & 0.1 & 0.6 & 0.3 & 140 & 120 & 0.0029 & 0.0010 \\
3C~256$^{(b)}$ & 1.4 & 9.2 & 7.0 & 1.0 & 0.6 & 0.3 & 3500 & 100 & 0.0192 & 0.0010 \\
3C~257 & 5.5 & 14.0 & 8.7 & 3.0 & 2.5 & 1.9 & 2500 & 300 & 0.0187 & 0.0010 \\
3C~266 & 6.2 & 2.6 & 1.5 & 0.8 & 0.3 & 0.2 & 500 & 100 & 0.0114 & 0.0010 \\
3C~297 & 8.5 & 3.0 & 1.9 & 2.4 & 0.6 & 0.5 & 900 & 350 & 0.0143 & 0.0010 \\
3C~305.1 & 5.2 & 4.2 & 3.0 & 1.1 & 0.3 & 0.2 & 700 & 120 & 0.0132 & 0.0010	\\
3C~324 & 3.2 & 5.5 & 3.5 & 1.4 & 0.5 & 0.3 & 1800 & 120 & 0.0174 & 0.0005 \\
3C~356 & 2.7 & 3.0 & 2.0 & 0.8 & 0.3 & 0.2 & 1000 & 140 & 0.0148 & 0.0005 \\
3C~368 & 6.4 & 5.4 & 3.1 & 1.1 & 0.6 & 0.3 & 600 & 100 & 0.0124 & 0.0010 \\
3C~454.1 & 13.0 & 9.1 & 7.2 & 4.1 & 1.3 & 1.6 & 1000 & 500 & 0.0148 & 0.0005 \\
3C~470 & 2.5 & 8.7 & 5.3 & 0.9 & 1.8 & 1.5 & 1200 & 350 & 0.0155 & 0.0005 \\
\hline 
\end{tabular}
\end{table*}
\subsection{The young stellar component}
When modelling the young stellar component (in the following interchangeably referred to as the starburst), we use the extreme case of an instantaneous burst of star formation \citep{RoccaVolmerange13}. Adopting such a formalism means that the starbursts in our work are by design in the post-starburst phase (star formation rate is equal to zero), where only the prescriptions from stellar evolution govern, among others, the production of dust grains and the overall SED shape. As expected from the robust {\herschel} detections of the objects in our sample, integrating these SEDs from 1 to 1000 $\mu$m yields IR starburst luminosities in excess of $10^{12}$ L$_{\odot}$ (Fig.~\ref{fig:all_vs_z}a), which puts them in the domain of ultra-luminous infrared galaxies \citep[ULIRGs,][]{Sanders&Mirabel96}.  

Table~\ref{tab:results} and Fig.~\ref{fig:all_vs_z}b show that the starbursts in the host galaxies of our 12 $z>1$ 3CR objects are massive ones, with estimated stellar masses well above 3 $\times$ 10$^{10}$ M$_{\odot}$ in all cases, and that they amount to $\sim$20 to $\sim$50\% of the total stellar mass of the systems. An exception to this trend is the host of 3C~470, wherein the young stellar component contains about 65\% of the total stellar mass. Given that 3C~470 is one of the objects with the highest redshifts, and the fact that the young stellar component amounts to about 50\% of the total stellar mass of the host of 3C~257 (redshift record-holder in the 3CR sample), these objects could also be supporting the view that their hosts are still going through the process of formation, with a significant amount of stellar mass yet to form. The stellar mass (and consequently the dust mass) of the starburst on average increases when going to high-redshift, which is explained in terms of the selection effect introduced by the {\herschel} detections.  

One of the main results from our work is the relatively evolved age of the starburst, which we uniformly find to be greater than 100 Myr across all objects (Fig.~\ref{fig:all_vs_z}c). While the uncertainty in the age estimates is at best a factor few, the best-fits strongly favour relatively evolved starbursts, allowing -- not surprisingly -- sufficient time for the process of stellar evolution to generate a dust content of order 10$^8$ M$_{\odot}$ (see Fig.~\ref{fig:all_vs_z}d). 
\subsection{The AGN component}
Recall that the high-$z$ 3CR RGs are powerful emitters in the mid-to-far infrared domain \citep[rest-frame wavelengths 5-40 $\mu$m][]{Haas08,Podigachoski15a}. The thermal continuum emission at these wavelengths is from the AGN-heated dust in the torus, which clearly dominates over the starburst emission following the diagnostics defined by \citet{Brandl06}. Similarly, considering the emission at 20 $\mu$m, which is the wavelength domain used by \citet{Drouart16} to link the AGN torus to the starburst component, we find that the torus emission is on average an order of magnitude stronger than that of the starburst at this wavelength domain. 

Quantifying the different physical parameters associated with the \citet{Siebenmorgen15} torus library is a challenging task given the small number of broadband MIR photometric data available, therefore we only address the torus luminosity which is calculated from the best-fit model as described above. Nevertheless, when considering all twelve objects studied here, we find that the viewing angles of the tori are always larger than 50$^\circ$ (consistent with these objects being radio galaxies), the cloud volume filling factors are generally large (about 80\%), and the inner radii of the tori are typically smaller than about 0.25 pc, which means that the sizes of the tori themselves are smaller than about 40 pc \citep{Siebenmorgen15}. Moreover, the optical depths of clouds avoid the extreme values in the associated parameter space, whereas the optical depths of the tori's disc midplane exhibit no preferential values. 

\begin{figure}
\includegraphics[width=\columnwidth]{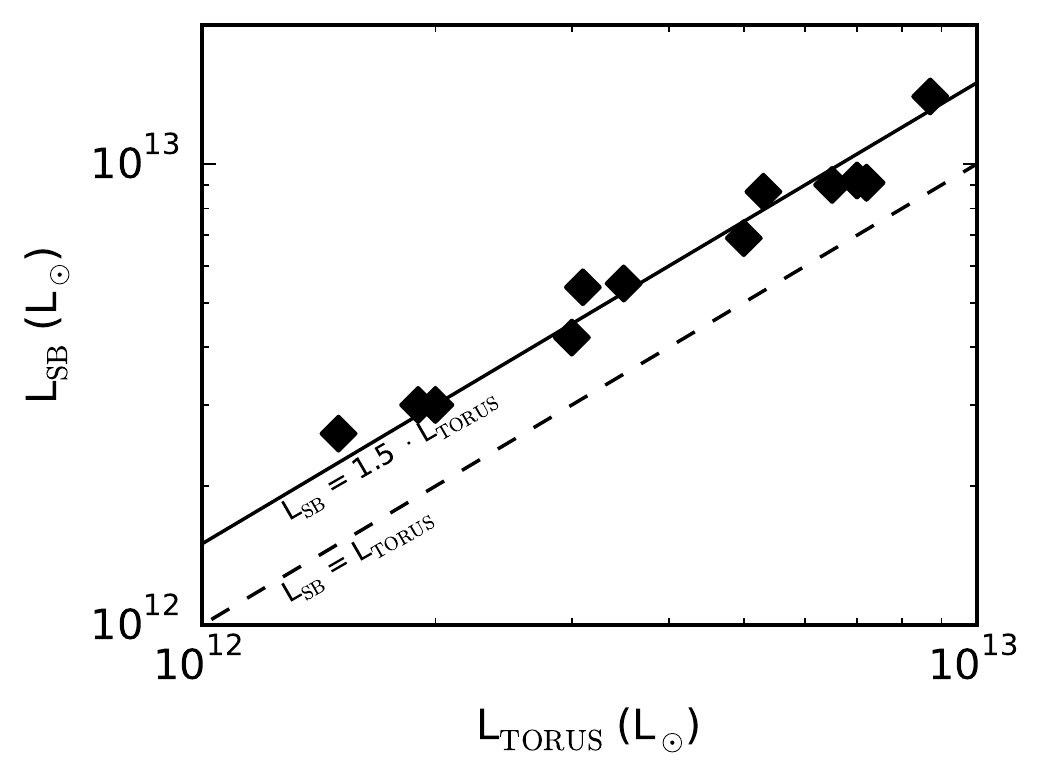}
\caption{Luminosity of the young stellar component as a function of that of the AGN torus component. The dashed and the solid lines indicate the one-to-one and the L$_{\mathrm{SB}}$ = 1.5 $\cdot$ L$_{\mathrm{TORUS}}$ lines, respectively.}
\label{fig:luminosities}
\end{figure}

The AGN torus luminosity is plotted in Fig.~\ref{fig:all_vs_z}a. Like the luminosity of the young stellar component, it ranges between $10^{12}$ and $10^{13}$ L$_{\odot}$, but is systematically lower than that of the young stellar component. Interestingly, the AGN torus luminosity appears to trace well the luminosity of the young stellar component across the entire redshift range, which is better visualized in Fig.~\ref{fig:luminosities}. The observed trend that higher starburst luminosities trace higher torus luminosities, based on the results obtained from the SED-fitting, is an intriguing one, and we return to this point in \S~\ref{sec:connection}.    
\subsection{The evolved stellar component}
\label{sec:esc}
The template representing the evolved stellar population is in many cases well constrained by the NIR and the shortest wavelength {\spitzer} data, often clearly showing the 1 $\mu$m peak due to evolved stars in the hosts of 3CR sources \citep{RoccaVolmerange13}. Hence, one of the more robustly estimated physical properties in this work is the stellar mass of the evolved stellar population. As shown in Fig.~\ref{fig:all_vs_z}b, this mass is of order $10^{11}$ M$_\odot$ and almost always dominates the stellar mass content of the 3CR host galaxies. While undoubtedly being massive, the 3CR RG hosts appear to be less massive than the most massive elliptical galaxies (10$^{12}$ M$_{\odot}$) corresponding to the brightest luminosity limit of the Hubble K-band diagram \citep{RoccaVolmerange04}. Using simple assumptions to subtract the young stellar and non-stellar contributions to only the optical and NIR photometry of the 3CR host galaxies studied in this work, \citet{Best98} estimated their stellar masses to be between a factor two to four more massive compared to our estimates. The availability of {\herschel} data which better constrain the contribution of young stars in the optical likely explains our somewhat lower stellar mass estimates compared to those by \citet{Best98}. A factor of four larger stellar masses for a few of our objects are also reported by \citet{Zirm03} and \citet{Targett11}, but these numbers are uncertain given that these studies are based exclusively on K-band photometry.

Less constrained is the age of this mature stellar population, which for about half of the objects in this work is in the range of a few billion years (see Fig.~\ref{fig:all_vs_z}c). The evolved stellar populations of these objects are consistent with being formed at high redshift and evolving passively to the redshift of observation. For the remaining objects, a somewhat lower age (between 0.5 and 1 Gyr) is preferred, and at least some of these are objects for which a contribution from an AGN component cannot be ruled out, so that the sum of an old and a young stellar component is not sufficient for a realistic fit of the observed SED in the UV/optical domain. An extreme example of this is 3C~256, which has both a stellar mass and age of the evolved population an order of magnitude lower than the corresponding properties of the other objects \citep[see details below and discussion by][]{Simpson99}. Another, more typical, example is 3C~368 (see Appendix for details), which is an object for which the likely AGN contribution leads to lower estimates of the age of the evolved stellar population. For such objects, the age of this stellar population should be treated as an absolute lower limit.   

The luminosity of the evolved stellar component is typically above $10^{11}$ L$_{\odot}$ and is on average an order of magnitude lower than that of either the AGN torus or the young stellar population. Most of the light in the 3CR radio galaxy hosts evidently is absorbed and re-emitted by dust at longer wavelengths.     
\section{Discussion}
\label{sec:discussion}

\subsection{AGN torus models}
\begin{figure}
\includegraphics[width=\columnwidth]{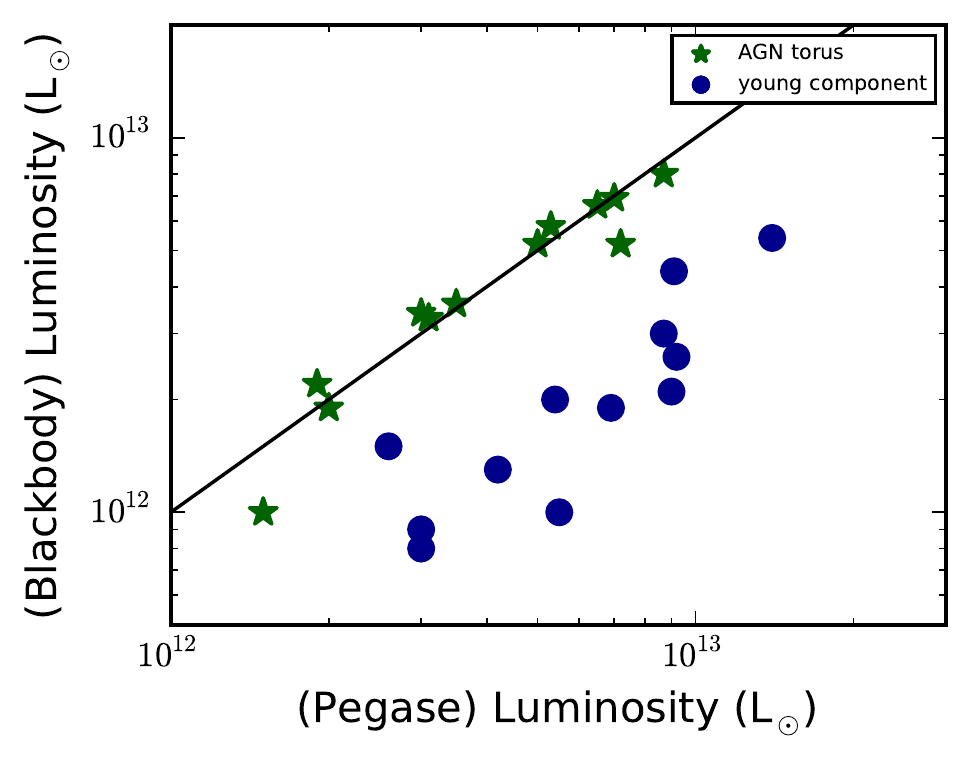}
\caption{Comparison between the luminosities calculated using the {\pegase}.3 templates and \citet{Siebenmorgen15} models (this work) and those calculated using a modified blackbody component and \citet{Hoenig&Kishimoto10} models \citep{Podigachoski15a}. The torus and young stellar components are plotted as (green) stars and (blue) circles, respectively. The one-to-one line is indicated with the solid line.}
\label{fig:comparison}
\end{figure}
Prior to this work, \citet{Drouart16} analysed the UV-to-submm SEDs of a sample of $1 < z < 4$ radio galaxy hosts using templates based on the {\pegase}.3 models, and the \citet{Fritz06} library of smooth torus models. We briefly compare the results obtained for two objects, 3C~368 and 3C~470, for which much of the same\footnote{The input {\herschel} flux densities are marginally different, whereby we use the photometry by \citet{Podigachoski15a} and \citet{Drouart16} used that by \citet{Drouart14}.} input data is used in both studies. Considering the best-fit SEDs of these two objects, \citet{Drouart16} found that the rest-frame wavelength domain between 10 and 50 $\mu$m is dominated by the starburst emission, whereas we find that the emission in this domain is largely dominated by the torus emission. This most striking difference is a direct consequence of the adoption of the \citet{Siebenmorgen15} models in the current work, which results in overall somewhat better fits to the {\herschel} photometry. We refrain from discussing the physical properties of the tori themselves, but we maintain that the \citet{Siebenmorgen15} models also provide excellent fits to the MIR photometric, and more importantly, spectroscopic observations of the 3CR radio galaxies (Fig.~\ref{fig:seds} and Fig.~\ref{fig:irs}). As a result, we find significantly higher torus luminosities and older starbursts, because the strong infrared torus emission associated with the clumpy models allows the peak of the starburst to move to longer wavelengths. The older starbursts in turn emit less extreme-UV radiation, allowing the old stellar components to account for more of the emission in this domain, hence resulting in younger ages compared to those estimated by \citet{Drouart16}. Evidently, the choice of the AGN torus models can greatly influence the results obtained from SED fitting \citep[but see][]{Feltre12}. A full comparison between the \citet{Drouart16} results and those presented here, taking into account the different samples (i.e., different redshift ranges, AGN luminosities, PAH-feature strengths, alignment effect strengths, etc.) is postponed for a future study; how much of the differences can be explained with the different torus models remains to be seen. 

\citet{Podigachoski15a} fitted the infrared SEDs of the complete sample of $z>1$ 3CR RGs (and quasars) with the sum of a clumpy torus model from \citet{Hoenig&Kishimoto10} and a modified blackbody component. These authors found that in all but two RGs, the AGN torus component has a higher infrared luminosity compared to that of the SF component. We compare these luminosities to those computed in our current work, and present the comparison in Fig.~\ref{fig:comparison}. On the one hand, the AGN torus luminosities in the two approaches are nearly identical, thereby confirming the reliability of the SED decomposition in the infrared domain. On the other hand, the starburst luminosities are systematically different, such that when using the modified blackbody approach, one underestimates this quantity by a factor of three on average. This difference, however, is easily reconciled when considering that the approach featuring a modified blackbody disregards the additional luminosity emitted by the aromatic features and the continuum radiation at the shorter (few micron) NIR/MIR wavelengths.  
\subsection{The starburst-torus connection}
\label{sec:connection}
\begin{figure}
\includegraphics[width=\columnwidth]{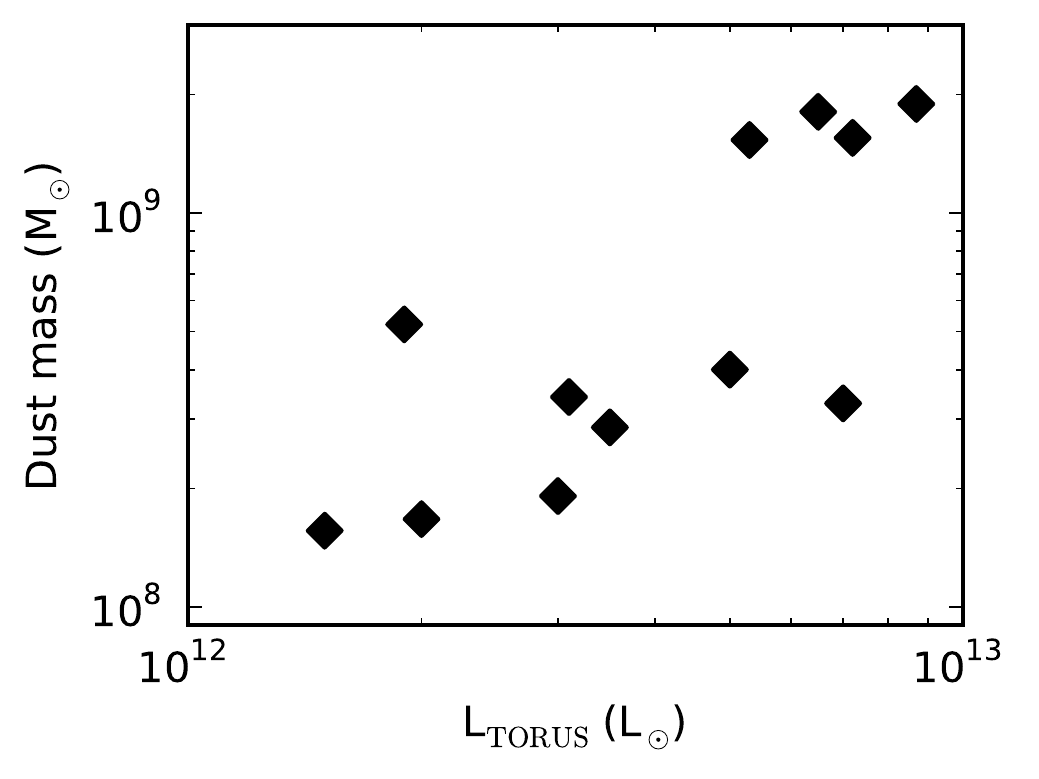}
\caption{Dust mass of the starburst component versus the infrared (1-1000$\mu$m) luminosity of the AGN torus component.} 
\label{fig:connection}
\end{figure}
The possible link between the black hole and star formation activity in distant galaxies is largely debated in the literature, with a number of studies involving {\herschel} observations reporting either no/weak or very strong correlation between these two processes \citep[see reviews by][]{Hickox14,Lutz14}. In general, authors who find strong correlations \citep[e.g.,][]{Netzer09,Bonfield11} suggest that such correlations arise naturally given that both processes are being fuelled by the same material, possibly of origins external to the AGN host galaxy. 

Figure~\ref{fig:luminosities} shows a remarkable correlation between the luminosity of the starburst and that of the AGN torus. Both luminosities roughly cover one order of magnitude, which is also the range for the radio luminosity shown in Fig.~\ref{fig:selection}. As indicated by the solid line in the figure, in all cases the starburst luminosity exceeds that of the AGN torus by a factor of 1.5, with small scatter. The strong correlation and the constant ratio between the luminosities suggest that the two processes know about and possibly influence each other. One possible way in which this physical link could be established is if part of the dust produced in the current starburst ends up in the AGN torus, so that it provides extra material for intercepting the incoming radiation from the accretion disk. A consequence of this interpretation is that for objects of similar intrinsic AGN luminosity, like the 3CR RGs in our sample, the torus luminosity increases with increasing the starburst dust mass, in line with what is shown in Fig.~\ref{fig:connection}. For this interpretation to hold, the starbursts reported in this work should ideally take place in the circumnuclear regions, i.e., on scales <1 kpc, which could observationally be tested making use of the superb capabilities of the ALMA observatory.   

While the strong L$_{\mathrm{SB}}$-L$_{\mathrm{TORUS}}$ correlation clearly holds for the twelve RGs of the present work\footnote{In the case of 3C~305.1 and 3C~454.1, whose observed spectral energy distributions might also be consistent with a sum of an evolved stellar and torus component, the correlation as presented would not necessarily hold.}, it likely breaks down for the {\herschel}-undetected $z>1$ 3CR RGs. Recall that all RGs from the complete high-$z$ 3CR sample are detected with all instruments on {\spitzer} \citep{Haas08}, while only the most starbursting ones (close to 40\%) are detected at the longest {\herschel} wavelengths \citep{Podigachoski15a} such that their young stellar components are as luminous as the AGN-tori. This means that powerful starbursts are associated with luminous torus components, but not the other way around. Hence, the correlation presented in Fig.~\ref{fig:luminosities} is likely the envelope of the L$_{\mathrm{SB}}$-L$_{\mathrm{TORUS}}$ relation, tracing the redshift-dependent luminosity, and the {\herschel}-undetected objects, with luminosities significantly below the ULIRG domain, likely populate regions well below this envelope.  
\subsection{Age effect}
\begin{figure}
\includegraphics[width=\columnwidth]{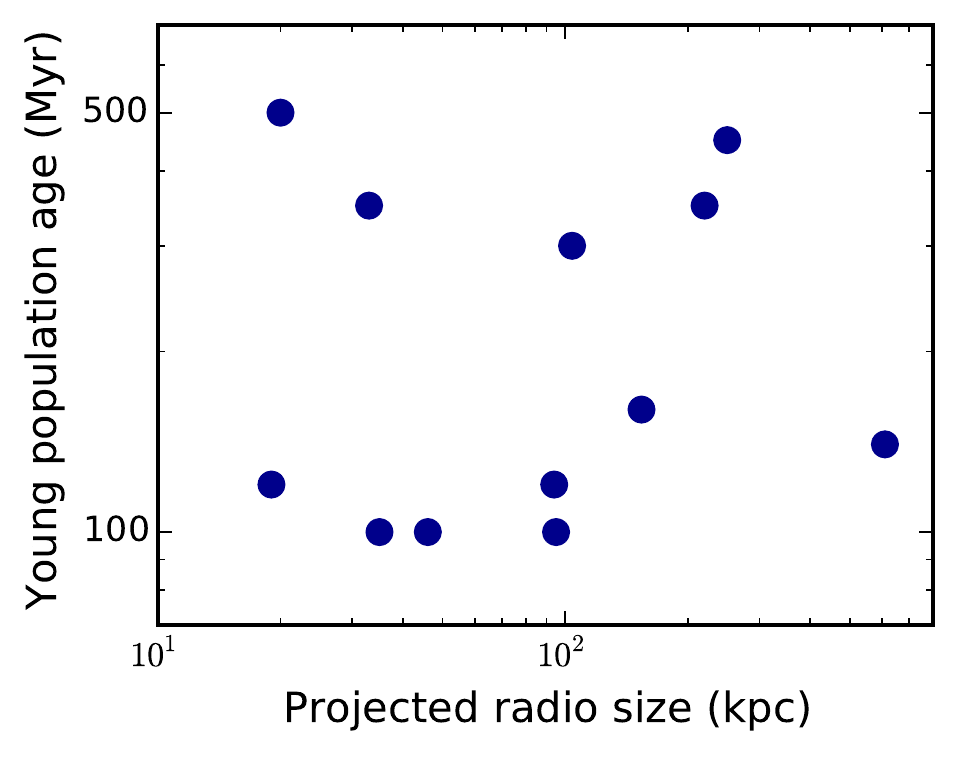}
\caption{Age of the young stellar population obtained from {\pegase}.3 as a function of the projected size of the radio source. The latter is the distance between the radio-lobes on either side of the AGN host, and is often used as a proxy for the age (duration) of the current AGN episode. }
\label{fig:ages_vs_sizes}
\end{figure}
High-resolution radio imaging provides a unique opportunity of measuring the lobe-to-lobe distance in RGs. Assuming a typical advance rate of the radio jet (10-20\% of the speed of light) and limited projection effects, such a measurement can be treated as a proxy for the duration of the current AGN episode. The radio size and the age of the young stellar population have been used in the literature to ascertain whether the triggering of the star formation and AGN activity are related to one another, and perhaps linked in an evolutionary scenario \citep[see e.g.,][and references therein]{Tadhunter11}. 

Figure~\ref{fig:ages_vs_sizes} shows the age of the young stellar population as a function of the projected size of the central radio source. No clear trend is evident, i.e., both small and large radio sources are found in hosts of both younger (100 Myrs) and somewhat older (300 - 500 Myr) young stellar populations. One interpretation of this result is that the onsets of the starburst and AGN activity do not follow a unique scenario, in line with what has been found by \citet{Tadhunter11} for nearby RGs. However, given the relatively evolved starbursts witnessed in our 3CR RG hosts, if both activities are triggered within a merger event \citep[e.g.,][]{Mihos&Hernquist96}, for which evidence was recently presented by \citet{Chiaberge15}, it appears more likely that the starbursts precede or are at least concurrent with the AGN episode. This interpretation, however, remains uncertain given the errors in the starburst age estimates. UV/optical spectra could in principle provide better age estimates, however the strong AGN activity and pronounced alignment effect in the high-$z$ 3CR population will likely complicate such age estimates. Based on the {\herschel} detection rates of their complete $z>1$ 3CR sample of RGs and quasars, \citet{Podigachoski15a} speculated that the starburst activity in the subgalactic-sized (<30 kpc) radio sources \citep[e.g.,][]{Odea16} may be triggered by the jet activity. The relatively evolved starbursts we find in the current work appear incompatible with this scenario of positive feedback, and the incidence (or lack) of feedback remains to be verified using the superb ALMA capabilities. 
\subsection{Caveat: apertures and non-stellar UV/optical continua}
\label{sec:limitations}
Simultaneously studying the properties of the stellar populations and the AGN-powered emission of statistically important samples of RGs requires the availability of good quality photometric data covering much of the electromagnetic spectrum. This necessarily leads to using data from instruments with significantly different spatial resolutions, as for example in the case of the \textit{HST} and {\herschel} ({\spire}) data, which means that different physical regions are explored particularly in high-redshift objects like those studied in this work. Despite the fact that the apertures used for measuring the fluxes used in our work are often different, the effects on the obtained results are expected to be minimal, given that the apertures are often big enough to include most of the light from the AGN host galaxy. 

Powerful RGs often show evidence for some non-stellar contribution to their UV/optical continuum. This may be either due to AGN-powered nebular continuum \citep[e.g.,][]{Dickson95}, and/or light scattered off dust particles present in the polar regions \citep[e.g.,][]{DiSeregoAlighieri94}. The fact that the non-stellar contribution is not addressed here is, arguably, the main limitation of the current work. When fitting the UV/optical SEDs of high-$z$ RGs, \citet{Best98} approximated the total contribution from both processes with a power-law with a fixed spectral index, however, constraining the contribution from each process (and hence the power-law index) is not possible with the UV/optical data available for our work. However, based on the alignment effect (as discussed in \S~\ref{sec:seds}), we expect a non-stellar contribution to the SEDs of only a few RGs studied (i.e., 3C~256, 3C~266, and 3C~368), and in those cases, the results obtained should be treated with care (see also \S~\ref{sec:esc}). 

\begin{figure}
\includegraphics[width=\columnwidth]{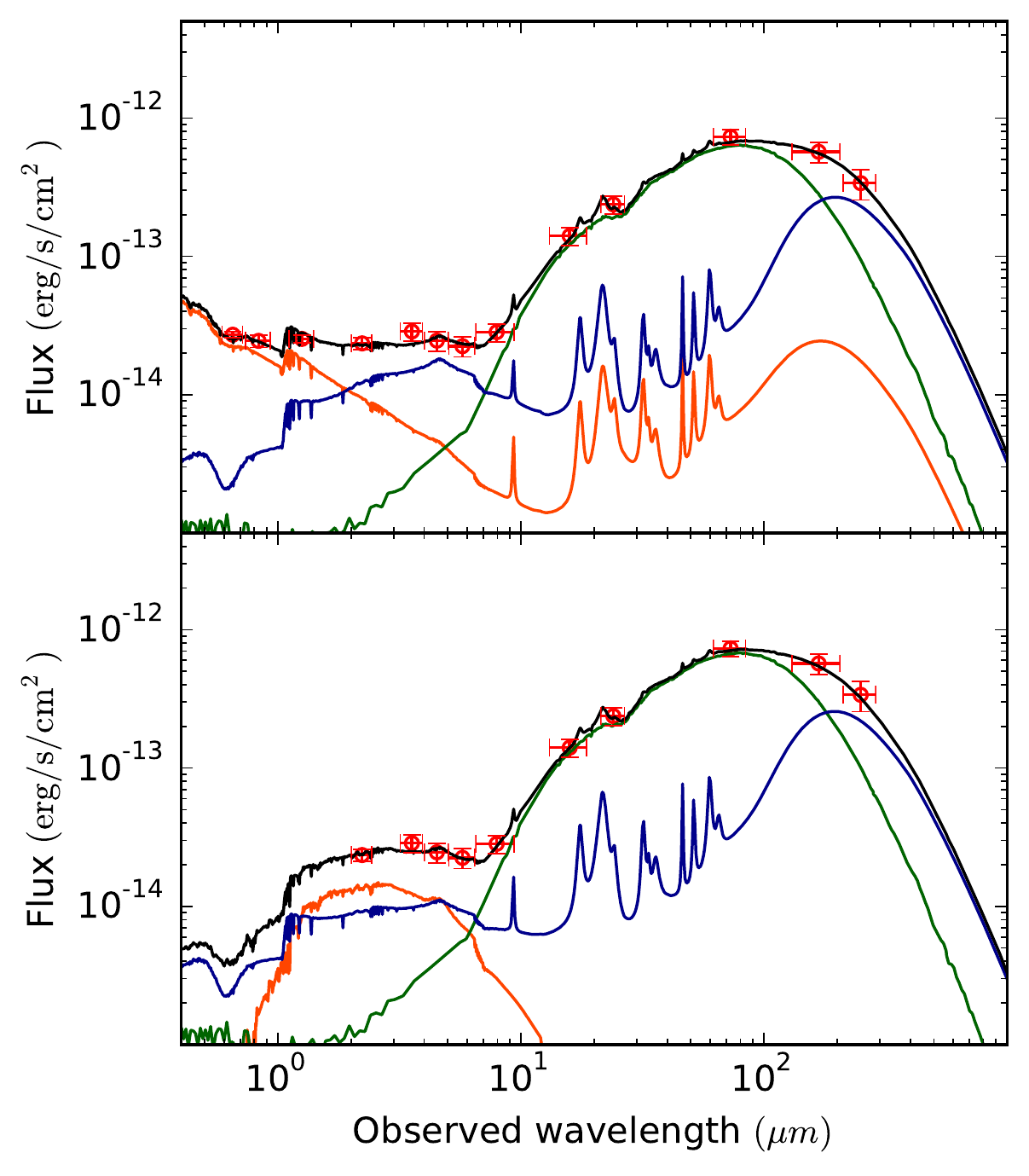}
\caption{Fitting the spectral energy distribution of 3C~256 using all available photometric data listed in Table~\ref{tab:3c256} (upper panel) and only data beyond K-band (lower panel). The lines and the symbols are identical as in Fig.~\ref{fig:seds}. While the physical properties estimated from the AGN and young stellar component are similar in both scenarios, those from the evolved stellar component are strikingly different (see text for more details), and clearly incorrect in the upper SED fit.}
\label{fig:256}
\end{figure}

The most striking example in which the spectral shape of the UV/optical SED is clearly of non-stellar origin is 3C~256 \citep{Simpson99}. A simple way to examine how the resulting physical properties are influenced by the presence of non-stellar contribution in 3C~256's host is to repeat the fit excluding the UV/optical photometric data (see Fig.~\ref{fig:256}). Comparing the two fits, the clear difference is in the region covered by the {\spitzer} IRAC bands, and without the requirement to fit the UV/optical points, the evolved stellar component contributes with equal weight compared to the young component in this domain. Furthermore, the extreme UV emission in the revised fit is dominated by the young stellar component. Not surprisingly, major differences are found in the estimated properties of the evolved stellar populations in the two scenarios: the stellar mass goes up by a factor of ten, and the age increases from 140 to 3000 Myr (see Table~\ref{tab:results}). Given that Fig.~\ref{fig:256} likely traces the extreme cases of old and young stellar component dominance in the UV domain, the results obtained for the old stellar component should be treated as limits. However, the AGN torus luminosity and the properties of the young stellar component (mass, age, luminosity, metallicity) remain within 10\% in both cases, which shows that the results obtained from the longer-wavelength part of the SED are fairly robust and not particularly sensitive on the possible presence of non-stellar contribution in the shortest wavelengths.  
\subsection{Concluding remarks}
We conclude the discussion section by presenting a number of possibilities for further exploration and verification of the results obtained in this work. One such possibility is to relax the extreme case of instantaneous starburst, and to investigate more plausible scenarios with somewhat longer duration of star formation activity. Another possibility is to extend our limited redshift coverage and break the redshift degeneracy by applying our approach, including the same torus models, to a larger compilation of objects consisting of the high-$z$ ($1 < z < 4$) RGs from the study by \citet{Drouart16} and the low-$z$ ones ($0.5 < z < 1$) addressed by \citet{Westhues16}. In such a study, two-component fits (evolved young population and a torus model) could also be considered for objects in which the starburst contributes only at the longest {\herschel} bands. Such a systematic study could also shed more light on the recently proposed idea of SMBH growth through accretion of supernova remnants accumulated along the star formation histories of RG hosts \citep{RoccaVolmerange15}.  
\section{Conclusions}
\label{sec:conclusions}
We studied the ultraviolet-to-submillimetre broad-band spectral energy distributions (SEDs) of a sample of twelve {\herschel}-detected 3CR radio galaxies covering the redshift range $1 < z < 2.5$, using stellar templates produced by the evolutionary code {\pegase}.3 and state-of-the-art templates describing the emission from AGN-heated dust in the torus. We find that the observed photometric SEDs, and in four objects also the {\spitzer} spectra, are in most cases well represented by a three-component model, which in addition to strong AGN torus emission includes both an evolved and massive stellar component (1 Gyr or older) and a starburst, confirming previous studies in the literature. The best-fit SEDs yield relatively evolved ($\sim$100 Myr or older) extremely massive starbursts which contain 20-50\% of the stellar mass of the systems, with infrared luminosities systematically larger than those of the AGN torus component. The observed correlation between these two luminosities is intriguing, but remains to be confirmed with a more robust analysis based on statistically larger samples of radio galaxy hosts.    
\section*{Acknowledgements}
This research was financially supported by Ammodo, through a Van Gogh travel grant. We thank Marco Chiaberge for kindly providing data prior to publication, and the expert referee for insightful comments and suggestions which improved the clarity of the paper. P.P. acknowledges the Nederlandse Organisatie voor Wetenschappelijk Onderzoek (NWO) for a Ph.D. fellowship, and the Leids Kerkhoven-Bosscha Fonds (LKBF) for a generous travel grant to the Institut d'Astrophysique de Paris.  
%
\bibliographystyle{mnras}
\bibliography{PEGASE3_Accepted}


\appendix
\section{Notes on individual spectral energy distributions}
%
%
\begin{table}
\centering
\caption{Table of photometric data used to fit the spectral energy distribution of 3C~068.2. The UV/optical data presented are emission-line corrected, as explained in the relevant references provided. Aperture sized used for the photometry can also be found in the same references. References: (1) \citet{Best97}, (2) \citet{Haas08}, (3) \citet{Podigachoski15a}.} 
\label{tab:3c068.2}
\begin{tabular}{cccr}
\hline
Band & Wavelength [$\mu$m] & Flux density [$\mu$Jy] & Ref.\\
\hline
HSTF785LP     &  0.9     &  7.9$\pm$0.9     &  (1)        \\ 
UKIRTJ        &  1.2     &  35.0$\pm$5.8    &  (1)        \\ 
UKIRTK        &  2.2     &  65.0$\pm$7.2    &  (1)        \\ 
IRAC1          &  3.6     &  105$\pm$16      &  (2)        \\ 
IRAC2          &  4.5     &  129$\pm$19      &  (2)        \\ 
IRAC3          &  5.8     &  137$\pm$21      &  (2)        \\ 
IRAC4          &  8.0     &  112$\pm$17      &  (2)        \\ 
IRS            &  16    &  1340$\pm$201      &  (2)        \\ 
MIPS1          &  24    &  1170$\pm$176      &  (2)        \\ 
PACS70        &  70    &  27500$\pm$2600    &  (3)        \\ 
PACS160       &  160   &  39600$\pm$5900    &  (3)        \\ 
SPIRE250      &  250   &  42000$\pm$7200    &  (3)        \\ 
SPIRE350      &  350   &  38700$\pm$7000    &  (3)        \\ 
SPIRE500      &  500   &  29000$\pm$7200    &  (3)        \\ 
\hline
\end{tabular}
\end{table}
\textit{3C~210} ---
\begin{table}
\centering
\caption{Same as Table~\ref{tab:3c068.2}, but for 3C~210. References: (1) \citet{Chiaberge15}, (2) \citet{Haas08}, (3) \citet{Podigachoski15a}.} 
\label{tab:3c210}
\begin{tabular}{cccr}
\hline
Band & Wavelength [$\mu$m] & Flux density [$\mu$Jy] & Ref.\\
\hline
HST3F606W     &  0.6     &  7.1$\pm$0.2     &  (1)        \\ 
HST3F140W     &  1.4     &  44.1$\pm$0.2    &  (1)        \\ 
IRAC1          &  3.6     &  256$\pm$38      &  (2)        \\ 
IRAC2          &  4.5     &  336$\pm$50      &  (2)        \\ 
IRAC3          &  5.8     &  489$\pm$73      &  (2)        \\ 
IRAC4          &  8.0     &  1090$\pm$164    &  (2)        \\ 
IRS            &  16    &  3410$\pm$512      &  (2)        \\ 
MIPS1          &  24    &  4430$\pm$665      &  (2)        \\ 
PACS70        &  70    &  31600$\pm$2400    &  (3)        \\ 
PACS160       &  160   &  56000$\pm$4000    &  (3)        \\ 
\hline
\end{tabular}
\end{table}
3C~210 has only two strong {\herschel} detections, however its redshift of $z=1.17$ ensures that the {\pacs} 160 $\mu m$ band probes emission beyond 70 $\mu$m, hence it is included in our sample. \\\\
\textit{3C~256} ---
\begin{table}
\centering
\caption{Same as Table~\ref{tab:3c068.2}, but for 3C~256. References: (1) \citet{Simpson99}, (2) \citet{Haas08}, (3) \citet{Podigachoski15a}.} 
\label{tab:3c256}
\begin{tabular}{cccr}
\hline
Band & Wavelength [$\mu$m] & Flux density [$\mu$Jy] & Ref.\\
\hline
CFHTR         &  0.7     &  5.9$\pm$0.3     &  (1)       \\ 
CFHTI         &  0.8     &  6.8$\pm$0.7     &  (1)       \\ 
NIRCJ         &  1.2     &  10.6$\pm$0.7     &  (1)       \\ 
NIRCK         &  2.2     &  17.4$\pm$1.7      &  (1)       \\ 
IRAC1         &  3.6     &  34$\pm$5      &  (2)          \\ 
IRAC2         &  4.5     &  37$\pm$6      &  (2)          \\ 
IRAC3         &  5.8     &  43$\pm$7      &  (2)          \\ 
IRAC4         &  8.0     &  75$\pm$11      &  (2)          \\ 
IRS           &  16    &  743$\pm$111      &  (2)          \\ 
MIPS1         &  24    &  1900$\pm$285      &  (2)          \\ 
PACS70        &  70    &  17800$\pm$2300      &  (3)  \\ 
PACS160       &  160   &  31900$\pm$5300      &  (3)  \\ 
SPIRE250      &  250   &  28200$\pm$6900      &  (3)  \\ 
\hline
\end{tabular}
\end{table}
This is one of the objects in which the alignment effect in the UV/optical is particularly pronounced \citep[see for example][]{Simpson99}. For a complete discussion on this issue, we refer the reader to \S~\ref{sec:limitations}, and in particular to Fig.~\ref{fig:256}. Note that by excluding the UV/optical photometric data in the revised fit in Fig.~\ref{fig:256}, 3C~256 technically no longer satisfies the selection criteria explained in \S~\ref{sec:data}. \\\\
\textit{3C~257} ---
\begin{table}
\centering
\caption{Same as Table~\ref{tab:3c068.2}, but for 3C~257. References: (1) \citet{Chiaberge15}, (2) \citet{Haas08}, (3) \citet{Podigachoski15a}, (4) \citet{Archibald01}.} 
\label{tab:3c257}
\begin{tabular}{cccr}
\hline
Band & Wavelength [$\mu$m] & Flux density [$\mu$Jy] & Ref.\\
\hline
HST3F606W     &  0.6     &  0.6$\pm$0.1    &  (1)        \\ 
HST3F140W     &  1.4     &  10.0$\pm$0.2     & (1)        \\ 
IRAC1          &  3.6     &  85$\pm$13      &  (2)          \\ 
IRAC2          &  4.5     &  111$\pm$17      &  (2)          \\ 
IRAC3          &  5.8     &  194$\pm$29      &  (2)          \\ 
IRAC4          &  8.0     &  322$\pm$48      &  (2)          \\ 
MIPS1          &  24    &  1360$\pm$204      &  (2)          \\ 
PACS70        &  70    &  8100$\pm$1000      &  (3)  \\ 
PACS160       &  160   &  15600$\pm$2500      &  (3)  \\ 
SPIRE250      &  250   &  33100$\pm$4700      &  (3)  \\ 
SPIRE350      &  350   &  31800$\pm$5500      &  (3)  \\ 
SPIRE500      &  500   &  32300$\pm$8600      &  (3)  \\ 
SCUBA850      &  850   &  5400$\pm$950      &  (4)     \\ 
\hline
\end{tabular}
\end{table}
This is the highest redshift ($z=2.47$) object in both our sample and in the complete $z>1$ 3CR radio-loud sample. \\\\
\textit{3C~266} ---
\begin{table}
\centering
\caption{Same as Table~\ref{tab:3c068.2}, but for 3C~266. References: (1) \citet{Best97}, (2) \citet{Haas08}, (3) \citet{Podigachoski15a}. The flux densities measured from the {\spitzer} IRS spectrum in the artificial broadband filters centred at 27, 30, and 33 $\mu$m are 1.7$\pm$0.3, 2.7$\pm$0.4, and 3.2$\pm$0.5 mJy, respectively.} 
\label{tab:3c266}
\begin{tabular}{cccr}
\hline
Band & Wavelength [$\mu$m] & Flux density [$\mu$Jy] & Ref.\\
\hline
HSTF555W      &  0.5     &  7.9$\pm$0.5     &  (1)         \\ 
HSTF702W      &  0.7     &  11.6$\pm$0.8     & (1)        \\ 
HSTF814W      &  0.8     &  15.7$\pm$0.5     &  (1)        \\ 
UKIRTK        &  2.2     &  56.1$\pm$4.7      &  (1)        \\ 
IRAC1          &  3.6     &  68$\pm$10      &  (2)          \\ 
IRAC2          &  4.5     &  73$\pm$11      &  (2)          \\ 
IRAC3          &  5.8     &  45$\pm$7      &  (2)          \\ 
IRAC4          &  8.0     &  102$\pm$15      &  (2)          \\ 
IRS            &  16    &  370$\pm$56      &  (2)          \\ 
MIPS1          &  24    &  980$\pm$147      &  (2)          \\ 
PACS70        &  70    &  7600$\pm$2400      &  (3)  \\ 
PACS160       &  160   &  29400$\pm$4100      & (3)  \\ 
SPIRE250      &  250   &  19500$\pm$5600      &  (3) \\
\hline
\end{tabular}
\end{table}
For this object, \citet{Inskip06} computed the UV/optical and near-infrared photometry using apertures of 4\arcsec in diameter \citep[as opposed to that of 9\arcsec used by][]{Best97}. We applied our fitting approach using the Inskip et al. data as input, and found that the estimated physical parameters remain practically the same (within 10\%). This is one of the four objects in the redshift range $1 < z < 1.4$ for which {\spitzer} IRS spectra are available. \\\\
%
%
\begin{table}
\centering
\caption{Same as Table~\ref{tab:3c068.2}, but for 3C~297. References: (1) \citet{Chiaberge15}, (2) \citet{Haas08}, (3) \citet{Podigachoski15a}.} 
\label{tab:3c297}
\begin{tabular}{cccr}
\hline
Band & Wavelength [$\mu$m] & Flux density [$\mu$Jy] & Ref.\\
\hline
HST3F606W     &  0.6     &  6.6$\pm$0.2     &  (1)        \\ 
HST3F140W     &  1.4     &  64.3$\pm$0.2     &  (1)        \\ 
IRAC1          &  3.6     &  119$\pm$18      &  (2)          \\ 
IRAC2          &  4.5     &  126$\pm$19      &  (2)          \\ 
IRAC3          &  5.8     &  122$\pm$18      &  (2)        \\ 
IRAC4          &  8.0     &  121$\pm$18      &  (2)          \\ 
MIPS1          &  24    &  432$\pm$65      &  (2)          \\ 
PACS70        &  70    &  12600$\pm$1200      &  (3)  \\ 
PACS160       &  160   &  15400$\pm$2400      &  (3)  \\ 
SPIRE250      &  250   &  24500$\pm$4300      &  (3)  \\ 
\hline
\end{tabular}
\end{table} \\\\
\textit{3C~305.1} ---
\begin{table}
\centering
\caption{Same as Table~\ref{tab:3c068.2}, but for 3C~305.1. References: (1) \citet{Chiaberge15}, (2) \citet{Haas08}, (3) \citet{Podigachoski15a}.} 
\label{tab:3c305.1}
\begin{tabular}{cccr}
\hline
Band & Wavelength [$\mu$m] & Flux density [$\mu$Jy] & Ref.\\
\hline
HST3F606W     &  0.6     &  10.7$\pm$0.2     &  (1)        \\ 
HST3F140W     &  1.4     &  60.3$\pm$0.2     &  (1)        \\ 
IRAC1          &  3.6     &  181$\pm$27      &  (2)        \\ 
IRAC2          &  4.5     &  282$\pm$42      &  (2)         \\ 
IRAC3          &  5.8     &  495$\pm$74      &  (2)        \\ 
IRAC4          &  8.0     &  972$\pm$146      &  (2)         \\ 
IRS            &  16    &  2410$\pm$362      &  (2)         \\ 
MIPS1          &  24    &  2490$\pm$374      &  (2)         \\ 
PACS70        &  70    &  24000$\pm$2300      &  (3)  \\ 
PACS160       &  160   &  40400$\pm$4300      &  (3)  \\ 
SPIRE250      &  250   &  34900$\pm$6000      &  (3)  \\ 
\hline
\end{tabular}
\end{table}
The UV-to-submm spectral energy distribution of this object could also be well represented with a two-component model, this being the sum of an evolved stellar component and the AGN torus component. In this case, the results for the evolved component would remain unchanged, whereas the luminosity of the torus would increase. \\\\ 
\textit{3C~324} ---
\begin{table}
\centering
\caption{Same as Table~\ref{tab:3c068.2}, but for 3C~324. References: (1) \citet{Chiaberge15}, (2) \citet{Best97}, (3) \citet{Haas08}, (4) \citet{Podigachoski15a}, (5) \citet{Best98_3c324}. The flux densities measured from the {\spitzer} IRS spectrum in the artificial broadband filters centred at 27, 30, and 33 $\mu$m are 6.3$\pm$1.0, 7.3$\pm$1.1, and 8.7$\pm$1.3 mJy, respectively.} 
\label{tab:3c324}
\begin{tabular}{cccr}
\hline
Band & Wavelength [$\mu$m] & Flux density [$\mu$Jy] & Ref.\\
\hline
HST3F606W     &  0.6     &  4.7$\pm$0.2     & (1)         \\ 
HSTF702W      &  0.7     &  7.8$\pm$0.6     & (2)         \\ 
HSTF791W      &  0.8     &  12.5$\pm$1.2      & (2)      \\ 
UKIRTJ        &  1.2     &  57.6$\pm$5.8      & (2)      \\ 
HST3F140W     &  1.4     &  48.3$\pm$0.2     & (1)        \\ 
UKIRTK        &  2.2     &  103.0$\pm$6.6      &  (2)        \\ 
IRAC1          &  3.6     &  165$\pm$25      &  (3)         \\ 
IRAC2          &  4.5     &  160$\pm$24      &  (3)          \\ 
IRAC3          &  5.8     &  178$\pm$27      &  (3)         \\ 
IRAC4          &  8.0     &  450$\pm$68      &  (3)        \\ 
IRS            &  16    &  2580$\pm$387     &  (3)         \\ 
MIPS1          &  24    &  2820$\pm$423     &  (3)         \\ 
PACS70        &  70    &  23500$\pm$2300      & (4)  \\ 
PACS160       &  160   &  31700$\pm$5600      & (4)  \\ 
SPIRE250      &  250   &  21000$\pm$6200      & (4) \\ 
SCUBA850      &  850   &  3010$\pm$720      & (5)      \\ 
\hline
\end{tabular}
\end{table}
This is one of the four objects in the redshift range $1 < z < 1.4$ for which {\spitzer} IRS spectra are available.\\\\
\textit{3C~356} ---
\begin{table}
\centering
\caption{Same as Table~\ref{tab:3c068.2}, but for 3C~356. References: (1) \citet{Chiaberge15}, (2) \citet{Best97}, (3) \citet{Haas08}, (4) \citet{Podigachoski15a}. The flux densities measured from the {\spitzer} IRS spectrum in the artificial broadband filters centred at 27, 30, and 33 $\mu$m are 5.7$\pm$0.7, 7.3$\pm$0.4, and 8.3$\pm$1.0 mJy, respectively.} 
\label{tab:3c356}
\begin{tabular}{cccr}
\hline
Band & Wavelength [$\mu$m] & Flux density [$\mu$Jy] & Ref.\\
\hline
HST3F606W     &  0.6     &  5.6$\pm$0.2     & (1)        \\ 
HSTF622W      &  0.6     &  7.4$\pm$0.6     & (2)        \\ 
HSTF814W      &  0.8     &  13.1$\pm$1.0      & (2)        \\ 
HST3F140W     &  1.4     &  45.3$\pm$0.2     &  (1)        \\ 
UKIRTK        &  2.2     &  64.4$\pm$3.6      & (2)        \\ 
IRAC1          &  3.6     &  108$\pm$16      & (3)          \\ 
IRAC2          &  4.5     &  110$\pm$16      &  (3)          \\ 
IRAC3          &  5.8     &  122$\pm$18      &  (3)          \\ 
IRAC4          &  8.0     &  434$\pm$65      &  (3)          \\ 
IRS            &  16    &  2270$\pm$341      &  (3)         \\ 
MIPS1          &  24    &  4060$\pm$609      &  (3)         \\ 
PACS70        &  70    &  11600$\pm$2500      & (4)  \\ 
PACS160       &  160   &  19700$\pm$4900      & (4) \\ 
\hline
\end{tabular}
\end{table}
3C~356 has only two strong {\herschel} detections, however its redshift of $z=1.08$ ensures that the {\pacs} 160 $\mu m$ band probes emission beyond 70 $\mu$m, hence it is included in our sample. This is one of the four objects in the redshift range $1 < z < 1.4$ for which {\spitzer} IRS spectra are available. \\\\
\textit{3C~368} ---
\begin{table}
\centering
\caption{Same as Table~\ref{tab:3c068.2}, but for 3C~368. References: (1) \citet{Best97}, (2) \citet{Haas08}, (3) \citet{Podigachoski15a}, (4) \citet{Archibald01}. The flux densities measured from the {\spitzer} IRS spectrum in the artificial broadband filters centred at 27, 30, and 33 $\mu$m are 6.4$\pm$1.0, 7.6$\pm$1.1 and 12.2$\pm$1.8 mJy, respectively.}
\label{tab:3c368}
\begin{tabular}{cccr}
\hline
Band & Wavelength [$\mu$m] & Flux density [$\mu$Jy] & Ref.\\
\hline
HSTF702W      &  0.7     &  21.7$\pm$2.6      & (1)        \\ 
HSTF791W      &  0.8     &  27.5$\pm$4.1      & (1)        \\ 
UKIRTJ        &  1.2     &  66.1$\pm$9.1      &  (1)       \\ 
UKIRTK        &  2.2     &  99.3$\pm$13.7      &  (1)     \\ 
IRAC1          &  3.6     &  126$\pm$19      & (2)          \\ 
IRAC2          &  4.5     &  112$\pm$17      & (2)         \\ 
IRAC3          &  5.8     &  112$\pm$17      &  (2)         \\ 
IRAC4          &  8.0     &  210$\pm$32      & (2)        \\ 
IRS            &  16    &  1370$\pm$206      &  (2)         \\ 
MIPS1          &  24    &  3250$\pm$488      &  (2)       \\ 
PACS70        &  70    &  29900$\pm$2000      &  (3) \\ 
PACS160       &  160   &  61500$\pm$4800      &  (3) \\ 
SPIRE250      &  250   &  44400$\pm$7400      &  (3)  \\ 
SPIRE350      &  350   &  23800$\pm$6200      &  (3)  \\ 
SCUBA850      &  850   &  4080$\pm$1080      &  (4)  \\ 
\hline
\end{tabular}
\end{table}
The optical/UV spectral shape requiring a relatively low age of the evolved stellar population, together with the shape of the NIR SED traced by the shorter {\spitzer} bands suggest that some AGN contribution is likely present in the optical photometry of this object. This might be consistent with \citet{Best98}, who argued that 3C~368 has the strongest alignment effect among the ones studied in their sample, so that the non-stellar contribution to the K-band photometry is about 30\%. This is one of the four objects in the redshift range $1 < z < 1.4$ for which {\spitzer} IRS spectra are available.\\\\
\textit{3C~454.1} ---
\begin{table}
\centering
\caption{Same as Table~\ref{tab:3c068.2}, but for 3C~454.1. References: (1) \citet{Chiaberge15}, (2) \citet{Haas08}, (3) \citet{Podigachoski15a}.} 
\label{tab:3c454.1}
\begin{tabular}{cccr}
\hline
Band & Wavelength [$\mu$m] & Flux density [$\mu$Jy] & Ref.\\
\hline
HST3F606W     &  0.6     &  3.6$\pm$0.4     & (1)         \\ 
HST3F140W     &  1.4     &  53.5$\pm$0.2     & (1)        \\ 
IRAC1          &  3.6     &  77$\pm$12      & (2)         \\ 
IRAC2          &  4.5     &  76$\pm$11      & (2)          \\ 
IRAC3          &  5.8     &  112$\pm$17      & (2)         \\ 
IRAC4          &  8.0     &  135$\pm$20      & (2)       \\ 
IRS            &  16    &  612$\pm$92      &  (2)       \\ 
MIPS1          &  24    &  1500$\pm$225      &  (2)         \\ 
PACS70        &  70    &  13700$\pm$2500      &  (3)  \\ 
PACS160       &  160   &  37000$\pm$4700      &  (3)  \\ 
SPIRE250      &  250   &  50200$\pm$8700      &  (3)  \\ 
SPIRE350      &  350   &  26700$\pm$10400      &  (3) \\ 
\hline
\end{tabular}
\end{table}
The UV-to-submm spectral energy distribution of this object might also be well represented with a two-component model, this being the sum of an evolved stellar component and the AGN torus component. In this case, the results for the evolved component would remain unchanged, whereas the luminosity of the torus would increase. \\\\
\textit{3C~470} ---
\begin{table}
\centering
\caption{Same as Table~\ref{tab:3c068.2}, but for 3C~470. References: (1) \citet{Best97}, (2) \citet{Haas08}, (3) \citet{Podigachoski15a}, (4) \citet{Archibald01}.} 
\label{tab:3c470}
\begin{tabular}{cccr}
\hline
Band & Wavelength [$\mu$m] & Flux density [$\mu$Jy] & Ref.\\
\hline
HSTF785LP     &  0.9     &  3.9$\pm$1.4      &  (1)         \\ 
UKIRTK        &  2.2     &  39.9$\pm$5.5      &  (1)         \\ 
IRAC1          &  3.6     &  50$\pm$7      &  (2)          \\ 
IRAC2          &  4.5     &  75$\pm$11     &  (2)          \\ 
IRAC3          &  5.8     &  72$\pm$11     &  (2)          \\ 
IRAC4          &  8.0     &  266$\pm$40     &  (2)         \\ 
IRS            &  16    &  1510$\pm$227      &  (2)       \\ 
MIPS1          &  24    &  2650$\pm$398      &  (2)         \\ 
PACS70        &  70    &  16000$\pm$2700      &  (3)  \\ 
PACS160       &  160   &  29300$\pm$5100      &  (3)  \\ 
SPIRE250      &  250   &  48000$\pm$6500      &  (3)  \\ 
SPIRE350      &  350   &  36300$\pm$5200      &  (3) \\ 
SCUBA850      &  850   &  5640$\pm$1080      &  (4)   \\ 
\hline
\end{tabular}
\end{table}
It is the only object in our sample, in which the stellar mass of the young stellar component is significantly larger than that of the evolved stellar component. \\\\\\
%
\bsp 
\label{lastpage}
\end{document}